\documentclass[aps,prd,reprint,showpacs,preprintnumbers,superscriptaddress,nofootinbib,floatfix]{revtex4-1}

\usepackage{graphicx,epsfig}
\usepackage{latexsym,amssymb,amsmath,float,url}
\usepackage[T1]{fontenc} 
\usepackage[ansinew]{inputenc}
\usepackage{epstopdf}
\usepackage{slashed}
\usepackage{bm} 
\usepackage{epsf}
  
\newcommand\nubar{{\bar\nu}}

\newcommand{\bea}{\begin{eqnarray}} 
\newcommand{\eea}{\end{eqnarray}} 
\newcommand{\nn}{\nonumber}

\mathchardef\mhyphen="2D 
 
\def\a{\alpha}

\def\m{\mu}

\def\p{\pi} 
 
\def\s{\sigma}

\def\x{\chi}

\newcommand{\half}{\frac{1}{2}}
\newcommand{\quarter}{\frac{1}{4}}

\newcommand{\eeq}{\end{equation}}

\newcommand{\barr}{\begin{array}}
\newcommand{\earr}{\end{array}}
\newcommand{\rarr}{\rightarrow}

\def\be{\begin{equation}}
\def\ee{\end{equation}}
\def\ba{\begin{eqnarray}}
\def\ea{\end{eqnarray}}
 
\begin{document}

\title{W/Z Bremsstrahlung as the Dominant Annihilation Channel for Dark Matter, Revisited}

\author{Nicole F.\ Bell} 
\affiliation{School of Physics, The University of Melbourne, Victoria 3010, Australia}

\author{James B.\ Dent}
\affiliation{Department of Physics and School of Earth and Space Exploration, Arizona State University, Tempe, AZ 85287-1404, USA}

\author{Ahmad J.\ Galea}
\affiliation{School of Physics, The University of Melbourne, Victoria 3010, Australia}

\author{Thomas D.\ Jacques}
\affiliation{School of Physics, The University of Melbourne, Victoria 3010, Australia}

\author{Lawrence M.\ Krauss}
\affiliation{Department of Physics and School of Earth and Space Exploration, Arizona State University, Tempe, AZ 85287-1404, USA}

\author{Thomas J.\ Weiler}
\affiliation{Department of Physics and Astronomy, Vanderbilt University, Nashville, TN 37235, USA}

\begin{abstract}
We revisit the calculation of electroweak bremsstrahlung contributions
to dark matter annihilation.  Dark matter annihilation to leptons is
necessarily accompanied by electroweak radiative corrections, in which
a $W$ or $Z$ boson is also radiated.  Significantly, while many dark
matter models feature a helicity suppressed annihilation rate to
fermions, bremsstrahlung process can remove this helicity suppression
such that the branching ratios Br($\ell \nu W $), Br($\ell^+\ell^-Z$),
and Br($\bar\nu \nu Z$) dominate over Br($\ell^+\ell^-$) and
Br($\bar\nu \nu$).  We find this is most significant in the limit where the dark matter mass is nearly degenerate with the mass of the boson which mediates the annihilation process.  Electroweak bremsstrahlung has important
phenomenological consequences both for the magnitude of the total dark
matter annihilation cross section and for the character of the
astrophysical signals for indirect detection.  Given that the $W$ and
$Z$ gauge bosons decay dominantly via hadronic channels, it is
impossible to produce final state leptons without accompanying
protons, antiprotons, and gamma rays.
\\
keywords: dark matter annihilation
\end{abstract}


\maketitle

\section{Introduction}
\label{sec:intro}

The importance of electroweak radiative corrections to dark matter
annihilation has recently been recognized, and examined in a number of
publications~\cite{Bell:2010ei,Berezinsky:2002hq,Kachelriess,BDJW,Dent:2008qy,KSS09,Ciafaloni:2010qr,Ciafaloni:2010ti,Bell:2011eu,marc}.
In a recent paper some of the present authors considered electroweak
bremsstrahlung contributions to dark matter annihilation, in models in
which dark matter annihilation to a fermion-antifermion pair,
$\chi\chi \rightarrow \bar{f}f$, is helicity
suppressed~\cite{Bell:2010ei}.  There it was shown that $W/Z$
bremsstrahlung lifts helicity suppressions, and can therefore be the
dominant DM annihilation mode.  However, some of
the quantitative conclusions of~\cite{Bell:2010ei} must be modified, as 
the explicit cross section calculation in therein was in error.
The purpose of the present paper is to revisit and extend the calculation of the $W/Z$
bremsstrahlung cross sections, and draw inferences  from the result.  
The main inference is that the three body final state processes can still dominate the tree level process
as claimed in~\cite{Bell:2010ei}.
We show herein that the claim finds support in the region where 
the parameter $\mu\equiv m_{\eta}^2/m_{\chi}^2$ is not too far from unity, 
with $m_{\eta}$ and $m_{\chi}$ being the mass of the boson which mediates 
the annihilation process and the dark matter mass, respectively.  
This region of parameter space is reminiscent of the co-annihilation region in standard supersymmetric (SUSY) scenarios, 
although the present work can also be applied to models which are not in the SUSY framework.

Let us parametrize the dark matter annihilation cross section in the usual
way, 
\begin{equation}
\sigma v=a+bv^2, 
\end{equation}
where the constant $a$ arises from $s$-wave annihilation while the
constant $b$ receives contributions from both $s$- and $p$-wave
channels. Since the dark matter velocity in a galactic halo today is
$v\sim10^{-3}c$, the $p$-wave term is strongly velocity suppressed.  In
order to have a large annihilation cross section in the Universe
today, it is desirable to have an unsuppressed $a$ ($s$-wave) term.
However, the $s$-wave annihilation of DM to a fermion-antifermion pair
is helicity suppressed in a number of important and popular models.
The most well known example is the annihilation of supersymmetric
neutralinos to a fermion-antifermion pair.  
The circumstances under which helicity supressions do or
do not arise were discussed in detail in Ref.~\cite{Bell:2010ei}.

It has long been know that bremsstrahlung of photons can lift such a
helicity suppression, leading to the result that the cross section for $\chi\chi
\rightarrow \bar{f} f \gamma$ can dominate over that for $\chi\chi
\rightarrow \bar{f}
f$~\cite{gamma1,gamma2,gamma3,gamma4,gamma5,gamma6}. However, the fact
that radiation of a $W$ or $Z$ gauge boson would also lift a helicity
suppression had been overlooked until the work of
Refs.~\cite{Bell:2010ei,BDJW}.  In these scenarios for which the
helicity suppression is removed, the dominant annihilation channels
are the set of bremsstralung processes, namely $\gamma$, $W$ and $Z$
bremsstrahlung.  (If the dark matter annihilates to colored fermions,
radiation of gluons would also contribute).
The phenomenology of $W$ and $Z$ bremsstrahlung is richer than that for
photon bremsstrahlung alone.  This is because the $W$ and $Z$ bosons
decay dominantly to hadronic final states, including antiprotons, for
which interesting cosmic ray bounds exist.

\section{Example of suppressed annihilation} 
\label{sec:model}

To illustrate our arguments, we choose a simple example of the class
of model under discussion.  This is provided by the leptophilic
model proposed in Ref.~\cite{Cao:2009yy,Ma:2000cc}.
Here the DM consists of a gauge-singlet Majorana fermion $\chi$ which
annihilates to leptons via the $SU(2)$-invariant interaction term
\begin{equation}
f\left(\nu\,\ell^-\right)_L\,\varepsilon\,
\left(
\barr{l}
\eta^+ \\
\eta^0 \\
\earr
\right)\chi + h.c.
= f(\nu_L\eta^0 - \ell^-_L \eta^+)\chi + h.c.
\label{eq:ma}
\end{equation}
where $f$ is a coupling constant, $\varepsilon$ is the $2\times 2$
antisymmetric matrix, and $(\eta^+$, $\eta^0)$ is a new $SU(2)$
doublet scalar.  In this model, DM annihilation to fermions is mediated
by $t$ and $u$ channel exchange of the $\eta$ fields.

An identical coupling occurs in supersymmetry if we identify $\chi$
with a neutralino and $\eta$ with a sfermion doublet.  In fact, the
implementation of supersymmetric photinos as dark matter by
H. Goldberg provided the first explicit calculation of $s$-wave
suppressed Majorana dark matter annihilation to a fermion pair~\cite{Haim1983}.  
Therefore, much of 
what we discuss below is also relevant for neutralino annihilation to
fermions via the exchange of sfermions.  However, the class of models
for which the $2\rightarrow 2$ annnihilation is helicity suppressed is
more general than the class of supersymmetric models.

The cross section for the $2\rarr 2$ process 
$\chi\chi\rightarrow e^+e^-$~or~$\nu\nubar$ is given by
\begin{eqnarray}
v\,\sigma = \frac{f^4 v^2 }{24\pi\,m_\chi^2}\,
\frac{1+\mu^2}{(1+\mu)^4}\,,
\label{eq:tree}
\end{eqnarray}
where $m_l\simeq0$ and $m_{\eta^\pm}=m_{\eta^0}$ have been assumed,
and $\mu=m_\eta^2/m_\chi^2$.
The suppressions discussed above are apparent in Eq.~(\ref{eq:tree}).
The helicity suppressed $s$-wave term is absent in the $m_l\ =0$
limit, and thus only the $v^2$-suppressed term remains.

\begin{figure*}[t]
\centering
\includegraphics[width=0.64\columnwidth]{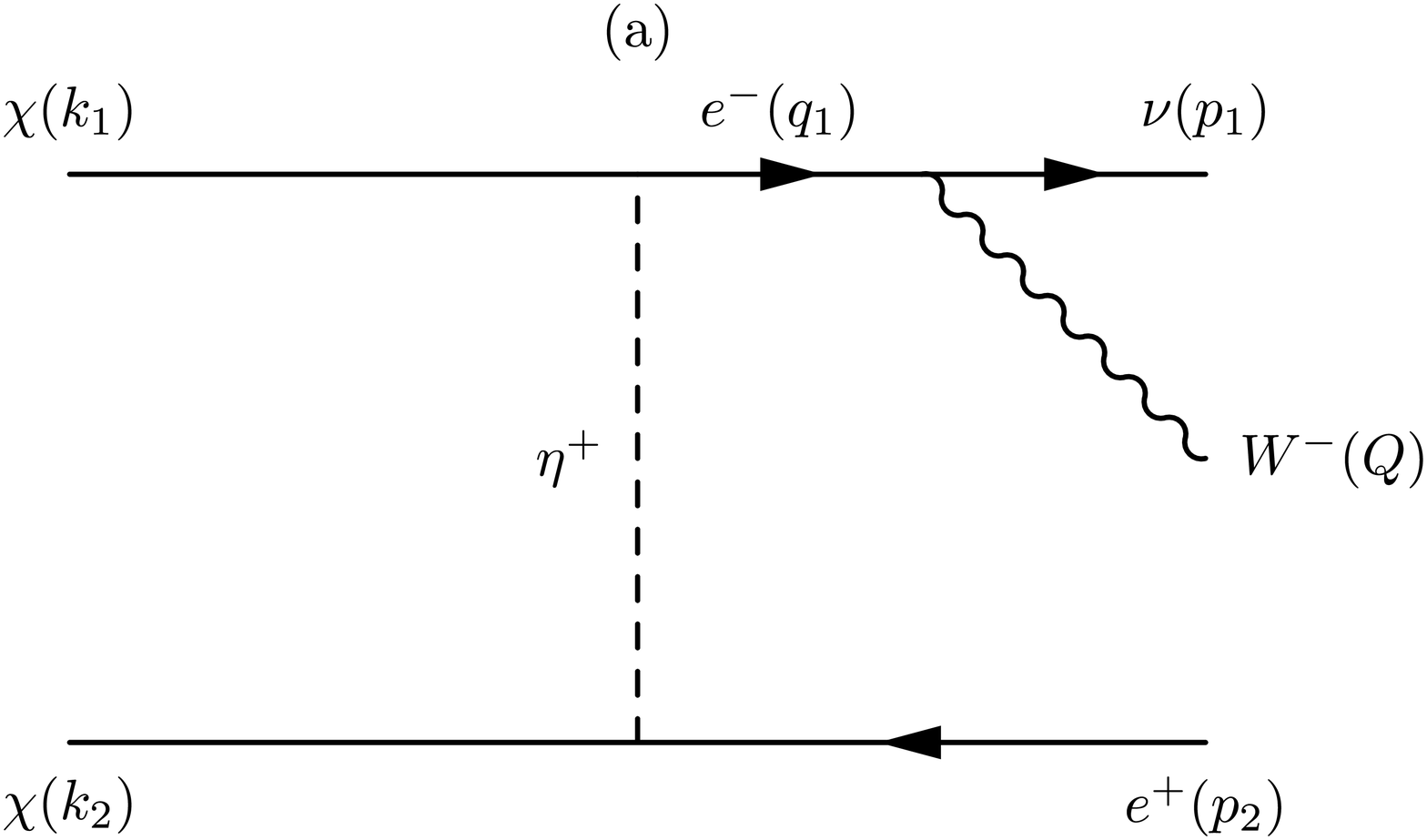}
\includegraphics[width=0.64\columnwidth]{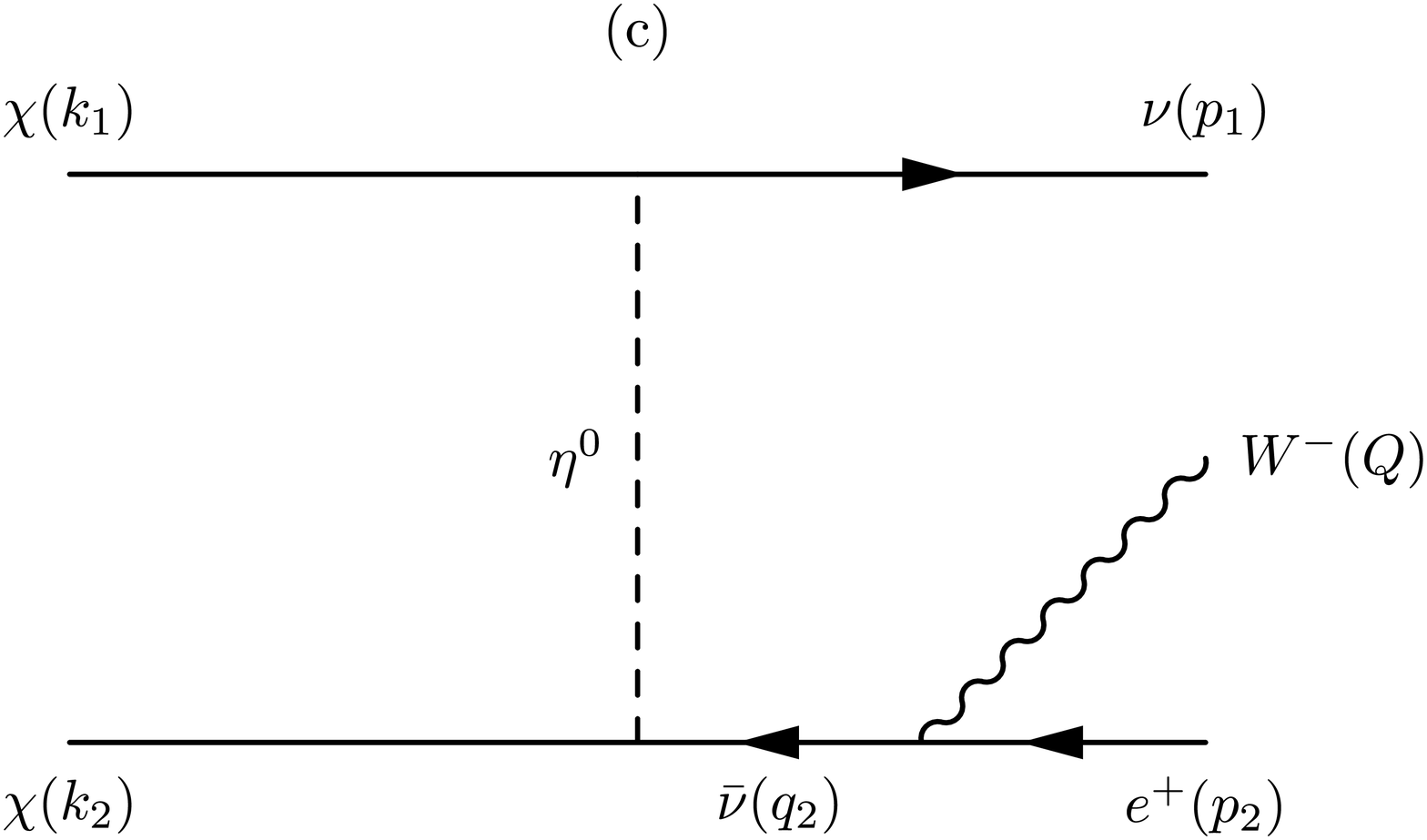}
\includegraphics[width=0.64\columnwidth]{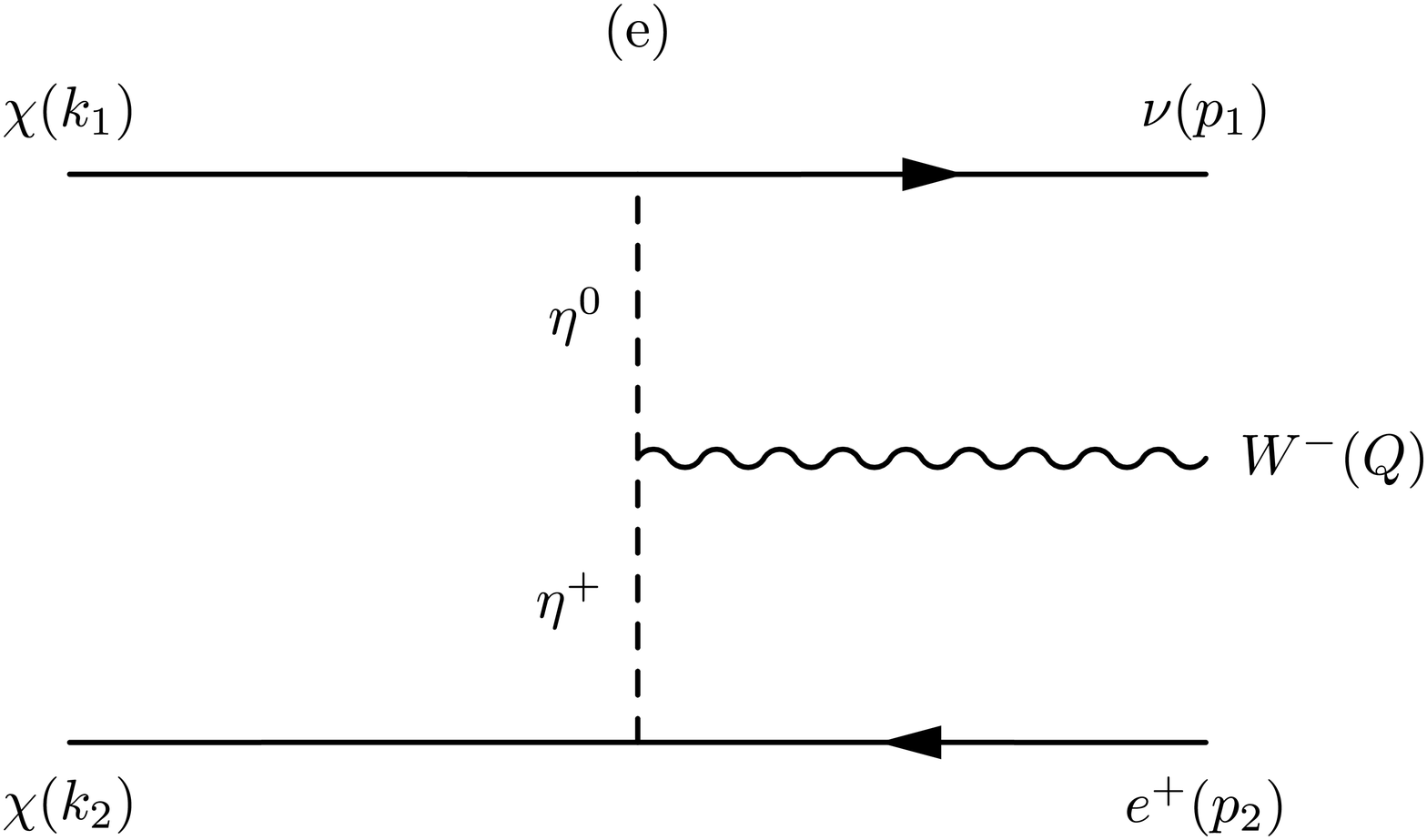}\\
\includegraphics[width=0.64\columnwidth]{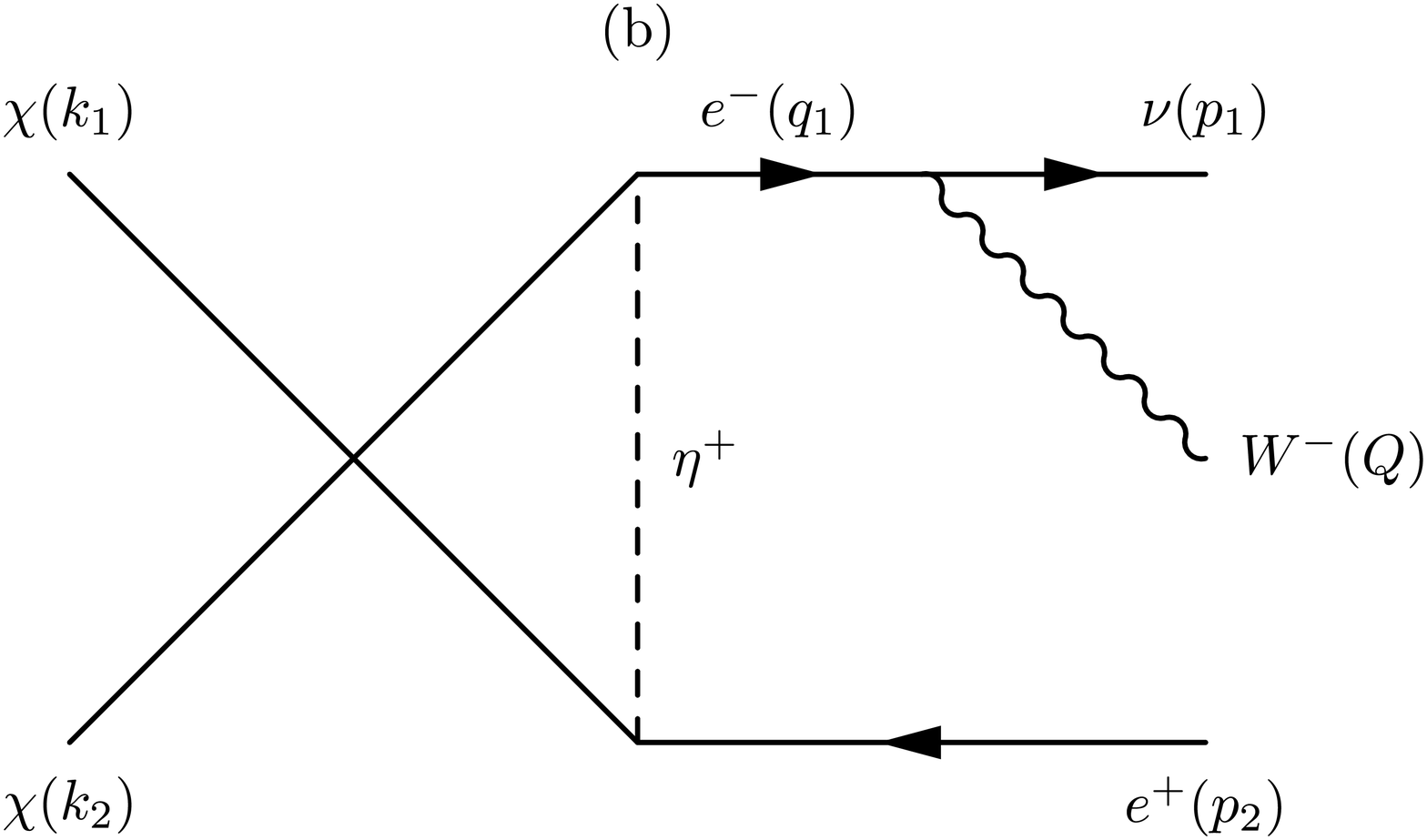}
\includegraphics[width=0.64\columnwidth]{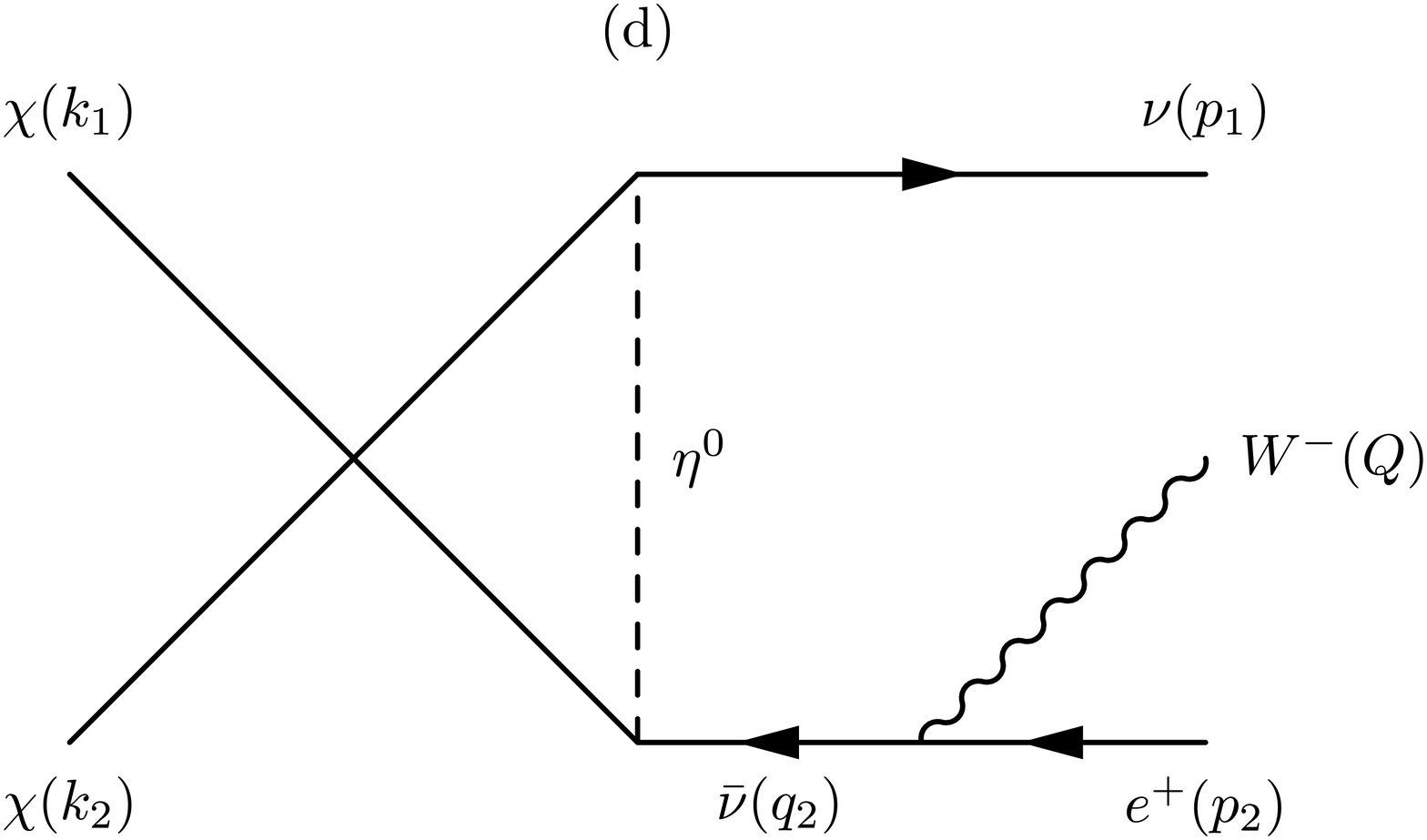}
\includegraphics[width=0.64\columnwidth]{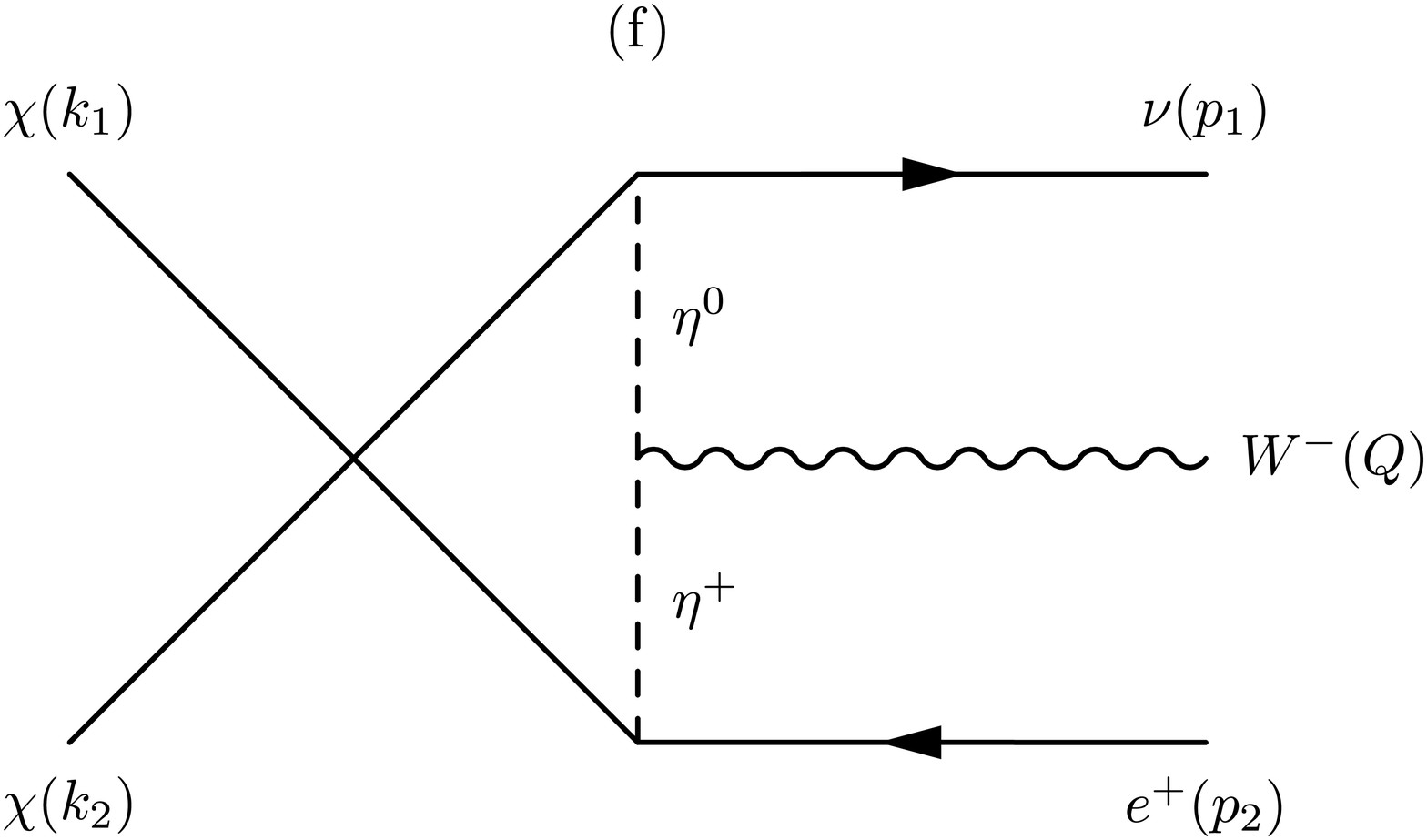}
\caption{
\label{fig:feyngraphs} 
The $t$-channel ((a),(c), and (e)) and $u$-channel ((b), (d) and (f))
Feynman diagrams for $\chi\chi\rightarrow e^+\nu W^-$.  
Note that $t$- and $u$-channel amplitudes are simply related by the 
$k_1\leftrightarrow k_2$ interchange symmetry.
All fermion momenta in the diagrams flow with the arrow except $p_2$ and $q_2$, 
with $q_1=p_1+Q$, $q_2=p_2+Q$.}
\end{figure*} 

\section{Lifting the suppression with electroweak bremsstrahlung} 
\label{sec:unsuppress}

\subsection{W-strahlung cross section}
\label{subsec:Wbrem}

We shall take the limit $m_l\simeq 0$ and assume that $m_{\eta^\pm}=m_{\eta^0}$.
The matrix elements for the six diagrams of Fig.~\ref{fig:feyngraphs}  are given by
\begin{eqnarray}
\label{matrixa}
\nonumber
\mathcal{M}_a &=&i \frac{f^2g}{\sqrt{2}}\frac{1}{q_1^2}\frac{1}{t_1-m_{\eta}^2} \nonumber\\
&&\times(\bar{v}(k_2)P_L v(p_2))(\bar{u}(p_1)\gamma^{\mu}P_L\slashed{q}_1u(k_1))\epsilon_{\mu}^Q,\\
\label{matrixb}
\mathcal{M}_b &=& i\frac{f^2g}{\sqrt{2}}\frac{1}{q_1^2}\frac{1}{u_1-m_{\eta}^2}\nonumber\\
&&\times(\bar{v}(k_1)P_L v(p_2))(\bar{u}(p_1)\gamma^{\mu}P_L\slashed{q}_1u(k_2))\epsilon_{\mu}^Q,\\
\label{matrixc}
\mathcal{M}_c &=& -i\frac{f^2g}{\sqrt{2}}\frac{1}{q_2^2}\frac{1}{t_2-m_{\eta}^2}\nonumber\\
&&\times(\bar{v}(k_2)P_L \slashed{q}_2\gamma^{\mu}v(p_2))(\bar{u}(p_1)P_Ru(k_1))\epsilon_{\mu}^Q, \\
\label{matrixd}
\mathcal{M}_d &=& -i\frac{f^2g}{\sqrt{2}}\frac{1}{q_2^2}\frac{1}{u_2-m_{\eta}^2}\nonumber\\
&&\times((\bar{v}(k_1)P_L \slashed{q}_2\gamma^{\mu}v(p_2))(\bar{u}(p_1)P_Ru(k_2))\epsilon_{\mu}^Q,
\end{eqnarray}
\begin{eqnarray}
\label{matrixe}
\mathcal{M}_e &=& -i\frac{f^2g}{\sqrt{2}}\frac{1}{t_3-m_{\eta}^2}\frac{1}{t_3'-m_{\eta}^2}\nonumber\\
&&\times((\bar{v}(k_2)P_Lv(p_2))(\bar{u}(p_1)P_Ru(k_1))\nonumber\\
&&\times((k_1-p_1)+(k_1-p_1-Q))^{\mu}\epsilon_{\mu}^Q,\\\nonumber
\label{matrixf}
\mathcal{M}_f &=& -i\frac{f^2g}{\sqrt{2}}\frac{1}{u_3-m_{\eta}^2}\frac{1}{u_3'-m_{\eta}^2}\nonumber\\
&&\times((\bar{v}(k_1)P_Lv(p_2))(\bar{u}(p_1)P_Ru(k_2))\nonumber\\
&&\times((k_2-p_1)+(k_2-p_1-Q))^{\mu}\epsilon_{\mu}^Q,
\end{eqnarray}
where we define the usual helicity projectors 
$P_{R/L}\equiv \half (1\pm\gamma_5 )$, 
and the Mandelstam variables
\begin{eqnarray}\nonumber
t_1 &=& (k_1-q_1)^2,\\\nonumber
t_2 &=& (k_1-p_1)^2 = t_3,\\\nonumber
u_1&=&(k_2-q_1)^2,\\\nonumber
u_2&=& (k_2-p_1)^2=u_3,\\\nonumber
t_3'&=&(k_2-p_2)^2 = (k_1-p_1-Q)^2,\\\nonumber
u_3'&=&(k_1-p_2)^2 = (k_2-p_1-Q)^2.
\end{eqnarray}
The vertex factors used in the matrix elements are as follows: the
$l\nu W$ vertex has an
$\frac{ig}{\sqrt{2}}\gamma^{\mu}P_L\epsilon_{\mu}^Q$, the $\chi\eta l$
vertex is $ifP_L$, and the coupling between the $W^-$ and the
$\eta^+-\eta^0$ is taken to be of the form $-ig(p+p')/\sqrt{2}$ from
Ref.~\cite{haberkane}.
Fierz transformed versions of these matrix elements,
and some insight gained from them, are collected in
Appendix~\ref{fierztransformed}.


We have explicitly checked the gauge invariance of our set of Feynman
diagrams.  Writing the matrix element as
\begin{equation}
\mathcal{M} =\mathcal{M}^\mu\epsilon^Q_\mu, 
\end{equation}
the Ward identity
\begin{equation}
\label{ward}
Q_\mu \mathcal{M}^\mu =0, 
\end{equation}
is satisfied for the sum of the diagrams.  The Ward identity takes the
same form as for photon bremsstrahlung provided we take the lepton
masses to be zero, since the axial vector current is conserved in this limit.  
Note that diagrams (a)+(c)+(e) form a gauge invariant subset,
as do (b)+(d)+(f).  The full amplitude is the sum of the partial amplitudes, properly weighted
by a minus sign when two fermions are interchanged. Thus we have
$\mathcal{M} = (\mathcal{M}_a + \mathcal{M}_c + \mathcal{M}_e)
- (\mathcal{M}_b + \mathcal{M}_d + \mathcal{M}_f)$.

In performing the sum over spins and polarizations, we note the
standard polarization sum,
\begin{equation}
\sum_\text{pol.} \epsilon^Q_\mu\epsilon^Q_\nu = - \left( g_{\mu\nu} - \frac{Q_\mu Q_\nu}{m_W^2} \right),
\label{polsum}
\end{equation}
can be replaced with $-g_{\mu\nu}$ alone.  The Ward identity of
Eq.(\ref{ward}) ensures the second term in Eq.(\ref{polsum}) does
not contribute once the contributions from all diagrams 
are summed (and squared).

In addition, we find that the longitudinal polarization of the $W$ also does not contribute to 
the $s$-wave amplitude, i.e.
\begin{equation}
\mathcal{M}^\mu\epsilon^Q_{L\,\mu} = 0\,.
\end{equation}
The $W$ boson behaves as a massive transverse photon,
with just two transverse polarizations contributing.  
As a consequence, our calculation of 
$W$ bremsstrahlung must reduce to the known results for photon bremsstrahlung in the 
$m_W\rightarrow 0$ limit, modulo coupling constants.  
Below we will show that this happens.

The thermally-averaged cross section is given by
\begin{equation}
v\,d\sigma = \frac{1}{2s} \int\frac{1}{4} \sum_\text{spin, pol.} |\mathcal{M}|^2 
 \, dLips^3
\label{dsigma1}
\end{equation}
where the $\quarter$ arises from averaging over the spins of the initial $\chi$ pair,
$v=\sqrt{1-\frac{4m_\chi^2}{s}}$ is the mean dark matter velocity,
as well as the dark matter single-particle velocity in the center of mass frame\footnote
{Informative discussions of the meaning of $v$ are given in~\cite{LL}, 
and the inclusion of thermal averaging is covered in~\cite{GelmGondo}.}, and $dLips^n$ represents $n$-body Lorentz invariant phase space.

We calculate the cross section for $W$ emission following the
procedure outlined above, 
with the integration over phase space performed according to the
method described in Ref.~\cite{Bell:2010ei}.
We expand in powers of the DM velocity, $v$, keeping only the leading
order ($v^0$) contribution.  As expected, we have an unsuppressed
cross section given by
\begin{widetext}
\bea
\s v\simeq\frac{\a_W f^4}{256\p^2m_\x^2}&&\left\{\,(\m+1)\left[\frac{\p^2}{6} 
-\ln^2\left(\frac{2m_\x^2(\m+1)}{4m_\x^2\m-m_W^2}\right)
-2\mbox{Li}_2\left(\frac{2m_\x^2(\m+1)-m_W^2}{4m_\x^2\m-m_W^2}\right) \right.\right. \nn\\
&&+ 2\mbox{Li}_2\left(\frac{m_W^2}{2m_\x^2(\m+1)}\right)-\mbox{Li}_2\left(\frac{m_W^2}{m_\x^2(\m+1)^2}\right)
-2\mbox{Li}_2\left(\frac{m_W^2(\m-1)}{2(m_\x^2(\m+1)^2-m_W^2)}\right)
\nn\\
&&+2\ln\left(\frac{4m_\x^2\m-m_W^2}{2m_\x^2(\m-1)}\right)\ln\left(1-\frac{m_W^2}{2m_\x^2(\m+1)}\right)
\left.+\ln\left(\frac{m_W^2(\m-1)^2}{4(m_\x^2(\m+1)^2-m_W^2)}\right)\ln\left(1-\frac{m_W^2}{m_\x^2(\m+1)^2}\right)\right]\nn\\
&&+\frac{(4\m+3)}{(\m+1)}
- \frac{m_W^2\left(4m_\x^2(\m+1)(4\m+3)-(m_W^2-4m_\x^2)(\m-3)\right)}{16m_\x^4(\m+1)^2}\nn\\
&&+\frac{m_W^2\left(4m_\x^4(\m+1)^4-2m_W^2m_\x^2(\m+1)(\m+3)-m_W^4(\m-1)\right)}{4m_\x^4(\m+1)^3\left(m_\x^2(\m+1)^2-m_W^2\right)}
\ln\left(\frac{m_W^2}{4m_\x^2}\right)
\nn\\
&&+\ln\left(\frac{2m_\x^2(\m-1)}{2m_\x^2(\m+1)-m_W^2}\right)\frac{(\m-1)\left(2m_\x^2(\m+1)-m_W^2\right)}{4m_\x^4(\m+1)^3(4m_\x^2\m-m_W^2)\left(m_\x^2(\m+1)^2-m_W^2\right)}\nn\\
&&\Biggl.\times\left(4m_\x^6(\m+1)^4(4\m+1)-m_\x^4m_W^2(\m+1)^2\left(3\m(\m+6)+7\right)+2m_\x^2m_W^4\left(\m(\m+4)+1\right)-m_W^6\right)\Biggr\}
\label{Wcrosssection}
\eea
%
where 
$\alpha_W \equiv g^2/(4\pi)$ .  The
Spence function (or ``dilogarithm'') is defined as ${\rm Li_2}
(z)\equiv -\int^z_0 \frac{d\zeta}{\zeta}\ln|1-\zeta| =
\sum_{k=1}^\infty \frac{z^k}{k^2}$.  

If we take the limit $m_W\rightarrow 0$ and replace $\alpha_W$ with
$2\alpha_\text{em}$, then Eq.~(\ref{Wcrosssection}) reproduces the cross
section for bremsstrahlung of photons, 
namely\footnote{Note that Eq.2. of Ref.~\cite{gamma5} is larger by an overall factor of two, and also has the opposite sign for the $(1+\mu)[...]$ term, while Eq.1. of Ref.~\cite{gamma5} is consistent with our results.}
%
\bea
\s v\simeq\frac{\a_\text{em} f^4}{128\p^2m_\x^2}&&\left\{\,(\m+1)\left[\frac{\p^2}{6} 
-\ln^2\left(\frac{\m+1}{2\m}\right)
-2\mbox{Li}_2\left(\frac{\m+1}{2\m}\right) \right]
+\frac{4\m+3}{\m+1} + \frac{4\m^2-3\m-1}{2\m}\ln\left(\frac{\m-1}{\m+1}\right) \right\}.
\label{gammacrosssection}
\eea
\end{widetext} 
The successful recovery of the photon bremsstrahlung result in the
massless $W$ limit provides a check\footnote
{A related work~\cite{Ciafaloni:2011sa} appeared on the arXiv nearly simultaneously with ours.
In this related work there appears analytic expressions for the $M_Z,\ M_W = 0$ 
limits of the cross-section which we calculate, thereby providing another  calculational check.
}
 on the rather complicated
expression for massive $W$ bremsstrahlung given above in
Eq.(\ref{Wcrosssection}).

\begin{figure}[t]
\includegraphics[width=1.0\columnwidth]{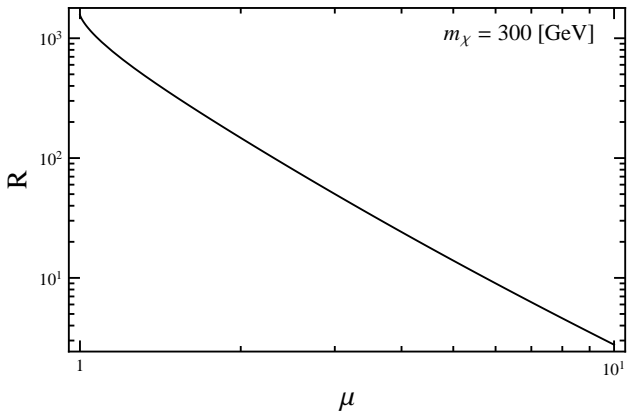}
\caption{
\label{ratio} 
The ratio $R=v\,\sigma (\chi\chi\to e^+ \nu W^-) /v\,\sigma
(\chi\chi\to e^+e^-)$ as a function of $\mu = (m_\eta/m_\chi)^2$, for
$m_\chi=300$ GeV.  We have used $v = 10^{-3}c$, appropriate for
the Galactic halo.}
\end{figure} 

\begin{figure}[t]
\includegraphics[width=1.0\columnwidth]{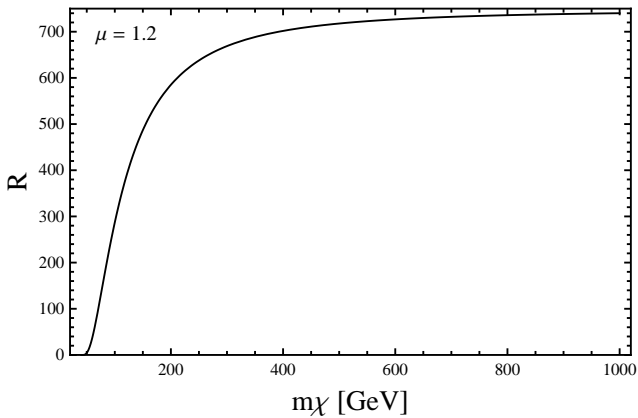}
\caption{
\label{ratio1} 
The ratio $R=v\,\sigma (\chi\chi\to e^+ \nu W^-) /v\,\sigma
(\chi\chi\to e^+e^-)$ as a function of the DM mass $m_\chi$, for
$\mu=1.2$ GeV.  We have used $v = 10^{-3}c$, appropriate for
the Galactic halo.}
\end{figure} 

\begin{figure}[t]
\includegraphics[width=1.0\columnwidth]{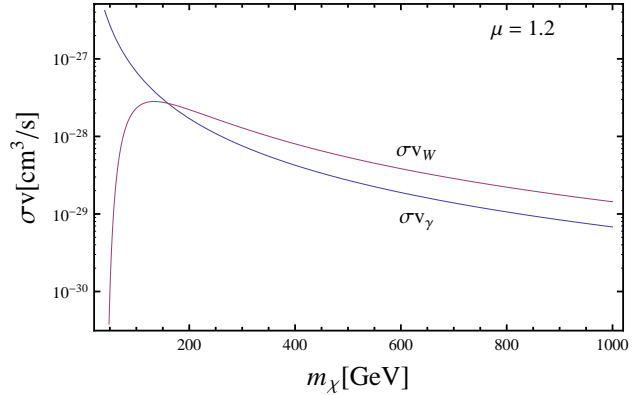}
\caption{
\label{sigma} 
The cross sections for $\chi\chi\to e^+ \nu W^-$ (red) and
$\chi\chi\to e^+ e^- \gamma$ (blue), for $\mu=1.2$ and coupling $f=1$.  
For large DM mass, the cross sections differ by a factor of $1/(2
\sin^2\theta_W)=2.17$ while for $m_\chi$ comparable to $m_W$ the W
bremsstrahlung cross section is suppressed by phase space effects.}
\end{figure} 

Since we are working in the limits $v=0$ and $m_f=0$, the nonzero
results in Eqs.(\ref{Wcrosssection}) and (\ref{gammacrosssection})
imply that the leading terms are neither helicity nor velocity
suppressed.  Not clear from the mathematical expressions is the
sensible fact that the cross sections fall monotonically with
increasing $m_\eta$ (or $\mu$).  This monotonic fall is shown in
Fig.~\ref{ratio}, where we plot the ratio of the $W$-strahlung cross
section to that of the lowest order process, $R = v\,\sigma
(\chi\chi\to e^+ \nu W^-) /v\,\sigma (\chi\chi\to e^+e^-)$.  The
lowest order process itself falls as $~ \mu^{-2}$, so the
$W$-strahlung process falls as $~\mu^{-4}$.  This latter dependence is
expected for processes with two propagators each off-shell by
$~1/\mu$, thereby signaling leading order cancellations among Fig. 1 diagrams (a)-(d).

Importantly, the effectiveness of the $W$-strahlung processes in
lifting suppression of the annihilation rate is evident in Fig.~\ref{ratio}.
The ratio is maximized for $\mu$ close to 1, where $m_\chi$ and $m_\eta$ are
nearly degenerate.  However, the $W$-strahlung process dominates over the
tree level annihilation even if a mild hierarchy between $m_\chi$ and
$m_\eta$ is assumed.  The ratio exceeds 100 for $\mu \alt 2$.

Fig.~\ref{ratio1} illustrates that the ratio $R$ is
insensitive to the DM mass, except for low $m_\chi$ where the $W$ mass
significantly impacts phase space.  From the figure one gleans that 
for $m_\chi\agt 3\,m_W$, the ratio $R$ is already near to its asymptotic value.
Incidentally, the asymptotic value may be obtained analytically by dividing 
Eq.~(\ref{gammacrosssection}) with Eq.~(\ref{eq:tree}) and 
rescaling $\alpha_{em}$ with $\alpha_W/2$.

In Fig.~\ref{sigma} we compare the $W$-strahlung cross section with
that for photon bremsstrahlung.  For high dark matter masses where the
$W$ mass is negligible, the two cross sections are identical except for
the overall normalization, which is higher by factor of $1/(2
\sin^2\theta_W)=2.17$ for $W$-strahlung.  For lower DM mass, the
available phase space is reduced due to $W$ mass effects, thus the
$W$-strahlung cross section falls below that for photons.  This can be
seen in Fig.~\ref{sigma} for $m_\chi \lesssim 150$ GeV (this number is
fairly insensitive to $\mu)$.  Another factor of two is gained for
$W$-strahlung when the $W^+$ mode is added to the $W^-$ mode shown
here.

Nominally, the correct dark matter energy fraction is obtained for
early-Universe thermal decoupling with an annihilation cross section
of $3\times 10^{-26} \textrm{cm}^3/\textrm{s}$.  It is seen in Fig.~\ref{sigma} that the
$W$-strahlung mode falls 2-3 orders of magnitude below this value.
Note that at the time of dark matter freeze-out in the early Universe,
the velocity suppression of the p-wave contribution is not as severe
as it is for late-Universe annihilation.  Hence, radiative
$W$-strahlung with its natural suppression factor $\alpha_W/4\pi$ is
probably not the dominant annihilation mode responsible for
early-Universe decoupling of Majorana dark matter.

\subsection{W and Lepton Spectra}
\label{subsec:spectra}
To obtain the energy spectrum of the $W$, we compute the differential
cross section in terms of $E_W$ by making the transformation
\begin{eqnarray}
d\cos(\theta_q) \rightarrow \frac{-4 \sqrt{s} q^2}{(s-q^2)(q^2-m_W^2)} dE_W. 
\end{eqnarray}
The energy spectrum of the the primary leptons is calculated in
similar fashion.
We find
\begin{widetext}
\bea
v \frac{d\s}{dx_W}= && \,\frac{\a_W f^4}{128\p^2m_\x^2}\,\left((1-x_W)+\frac{m_W^2}{4m_\x^2}\right)\left\{\sqrt{x_W^2-\frac{m_W^2}{m_\x^2}}\left[\frac{2}{\left((\m+1)(\m+1-2x_W)+\frac{m_W^2}{ m_\x^2}\right)}-\frac{1}{(\m+1-x_W)^2}\right]\right.\nn\\
&& \left. -\frac{\left((\m+1)(\m+1-2x_W)+\frac{m_W^2}{m_\x^2}\right)}{2(\m+1-x_W)^3}\ln\left(\frac{\m+1-x_W+\sqrt{x_W^2-m_W^2/m_\x^2}}{\m+1-x_W-\sqrt{x_W^2-m_W^2/m_\x^2}}\right)\right\},
\eea
\bea
v \frac{d\s}{dx_l}= &&\,\frac{\a_W f^4}{512\p^2m_\x^2}\frac{1}{(\m-1+2x_l)^2}\Biggl\{\left(4(1-x_l)^2-4x_l(\m+1)+3(\m+1)^2-\frac{m_W^2}{m_\x^2}(\m+3)\right)\Biggr.\nn\\
&&\times\ln\left(\frac{2m_\x^2(\m+1)(1-x_l)-m_W^2}{\left(2m_\x^2(\m+1-2x_l)-m_W^2\right)(1-x_l)}\right)-\frac{x_l\left(4m_\x^2(1-x_l)-m_W^2\right)}{\left(2m_\x^2(1-x_l)(\m+1)-m_W^2\right)(1-x_l)^2}\nn\\
&&\Biggl.\times\left[(1-x_l)^2\left(4(1-x_l)^2-x_l(\m+1)+3(\m+1)^2\right)+\frac{m_W^2}{4m_\x^2}(1-x_l)\left(x_l(\m+11)-4(\m+3)\right)-x_l\frac{m_W^2}{8m_\x^2}\right]\Biggr\}.
\eea
\end{widetext}

\begin{figure}[t]
\includegraphics[width=1.0\columnwidth]{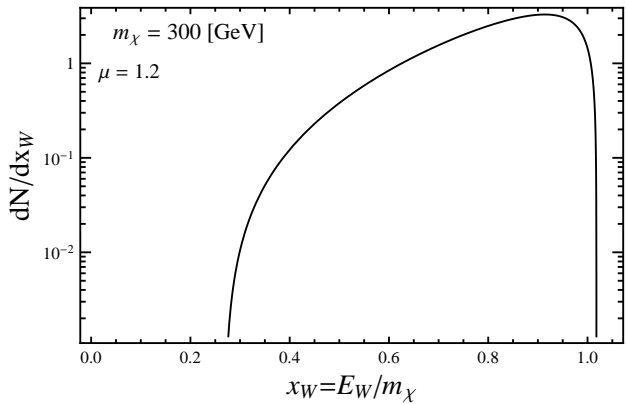}
\caption{
\label{dNdEW}
The $W$ spectrum per $\chi\chi\to e \nu W$ annihilation 
for $m_\chi=300$ GeV and $\mu = 1.2$.
}
\end{figure}

The $W$ spectrum per $\chi\chi\to e \nu W$ event is given in
Fig.~\ref{dNdEW}.  We use the scaling variable $x_W\equiv E_W/m_\chi$,
and plot $dN/dx_W \equiv (\frac{1}{\sigma_{e^+\nu W^-}})
\frac{d\sigma_{e^+\nu W^-}}{dx_W}$.  The kinematic range of $x_W$ is
$[\frac{m_W}{m_\chi},\,(1+\frac{m_W^2}{4m_\chi^2}) ]$, with the lower
limit corresponding to a $W$ produced at rest, and the upper limit
corresponding to parallel lepton momenta balancing the opposite $W$
momentum.  As evident in Fig.~\ref{dNdEW}, the $W$ boson spectrum has
a broad energy distribution, including a significant high energy
component.

\begin{figure}[t]
\includegraphics[width=1.0\columnwidth]{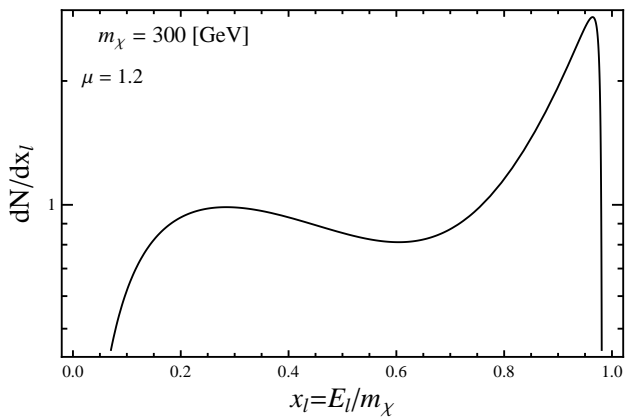}
\caption{
\label{dNdEe}
The primary lepton spectrum per $\chi\chi\to e \nu W$ annihilation, 
for $m_\chi=300$ GeV and $\mu = 1.2$.
}
\end{figure}

For the lepton energy spectrum, shown in Fig.~\ref{dNdEe}, the range
of the scaling variable $x_\ell\equiv E_\ell/m_\chi$ is
$[\,0,\,1-\frac{m_W^2}{4m_\chi^2} ]$.  Both limits arise when one
lepton has zero energy and the other is produced back-to-back with the
$W$.
Note that this spectrum is valid for either the
$e^+$ or the $\nu$ from the annihilation $\chi\chi\to e^+ \nu W^-$,
and for either $e^-$ or $\nubar$ from the annihilation $\chi\chi\to
e^- \nubar W^+$.

\subsection{Z Emission}

Consider the process producing the $\nubar\nu Z$ final state.  The
cross sections for the $Z$-strahlung processes are related to those for
$W$-strahlung in a simple way: The amplitudes producing $\nubar\nu Z$
arise from the same six graphs of Fig.\ref{fig:feyngraphs}, where $e$,
$W$ and $\eta^+$ are replaced everywhere by $\nu$ and $Z$ and
$\eta_0$, respectively.  The calculation of the amplitudes, and their
interferences, proceeds in an identical fashion.  After making
the replacement $m_W \rightarrow m_Z$, the cross section for the
annihilation process $\chi\chi\rightarrow\nu\bar \nu Z$ differs from
that for $\chi\chi\rightarrow e^+\nu W^-$ by only an overall
normalization factor,
\begin{eqnarray}
v\,\sigma_{\nu\bar\nu Z} &=& \left.\frac{1}{ (2\cos^2\theta_W )} \times
v\,\sigma_{e^+\nu W^-}\right|_{m_W\rightarrow m_Z} \nonumber\\ &\simeq&
0.65 \times v\,\sigma_{e^+\nu W^-} \Big|_{m_W\rightarrow m_Z} .
\end{eqnarray}

Consider now the $e^+e^-Z$ final state.  Again, the amplitudes arise
from the same six basic graphs of Fig.\ref{fig:feyngraphs}.  Since only
the left-handed leptons couple to the dark matter via the SU(2)
doublet $\eta$, only the left handed component of $e^-$ participates
in the interaction with the $Z$.  Therefore, the couplings of the
charged leptons to $Z$ and $W$
take the same form, up to a
normalization constant.  We thus find
\begin{eqnarray}
v\,\sigma_{e^+e^-Z}&=& 
\frac{2\left(\sin^2\theta_W - \frac{1}{2}\right)^2}{\cos^2\theta_W}
\times
v\,\sigma_{e^+\nu W^-} \, \Big|_{m_W\rightarrow m_Z}\nonumber\\
& \simeq & 0.19 \times \,v\,\sigma_{e^+\nu W^-}\Big|_{m_W\rightarrow m_Z}.
\end{eqnarray}
Adding the four contributions to $W/Z$strahlung, we find 
\begin{equation}
v\,\sigma_{W/Z-\textrm{strahlung}} = 2.84 \times v\,\sigma_{e^+\nu W^-}\,.
\end{equation}
%


\section{Discussion and Conclusions} 
\label{discussion}

There are clear advantages and disadvantages of seeking photon- versus
$W/Z$-bremsstrahlung as an indirect signature of dark matter.  With
photon bremsstrahlung, the photon itself is easily detected.  It's
energy spectrum may then be readily compared to model predictions.
With $W$-strahlung, it is the decay products of the $W$ decay which
must be sought.  Their spectra are less attributable to the model of
dark matter annihilation.  However, the total rate for $W/Z$-strahlung
exceeds that of photon-strahlung.  Photons couple with strength $e$,
$W$'s couple with strength $g/\sqrt{2} = e/(\sqrt{2}\sin\theta_W)$,
and $Z$'s couple to neutrinos with strength $g/(2\cos\theta_W) =
e/(2\cos\theta_W\sin\theta_W)$.  Therefore in the high energy limit
where the $W$ and $Z$ masses can be neglected, we expect
\bea
\sigma_{e^+\nu W^-} &=& \frac{1}{2 \sin^2\theta_W} \sigma_{e^+e^-\gamma}
= 2.17 \sigma_{e^+e^-\gamma}.
\eea
So, in the high energy limit where $m_\chi\agt 300 {\rm GeV} >> m_W$, 
the total cross section becomes
\bea
\sigma_\text{brem, total} &=& \sigma_{e^+\nu W^-} + \sigma_{\bar\nu e^- W^+}\nonumber\\
&&+\sigma_{\bar\nu \nu Z}+ \sigma_{e^+e^-Z} + \sigma_{e^+e^-\gamma}\nonumber\\
&=& 7.16\,\sigma_{e^+e^-\gamma}.
\eea

Furthermore, the varied decay products of the $W/Z$ allow more
multi-messenger experiments to engage in the dark matter search.
Charged leptons, protons and antiprotons, neutrinos, and even
deuterons are expected, at calculable rates and with predictable
spectra.  Importantly, hadronic decay products are unavoidable,
despite a purely leptonic tree-level annihilation.  The tens of
millions of $Z$ events produced at CERN's $e^+ e^-$ collider show in
detail what the branching fractions and spectra are for each kind of
decay product.  
In a forthcoming article~\cite{Bell:2011eu} we explore
the favorable prospects for using $W$-strahlung decay products as
indirect signatures for dark matter.

The lifting of the helicity suppression is most significant in the
limit where the mass of the boson mediating dark matter annihilation
does not greatly exceed the mass of the dark matter particle.  This
is true both for photon bremsstrahling and for $W$/$Z$-bremsstrahlung.
In this limit, we find the three body final state annihilation
channels can significantly dominate over two body annihilation
channels.  The region of parameter space where $\chi$ and $\eta$ are
approximately degenerate is of great interest in many models, since it
coincides with the co-annihilation region where both $\chi\chi$ and
$\chi\eta$ annihilations are important in determining the relic dark
matter density at the time of freezeout in the early Universe, often a
favored parameter region in SUSY scenarios.

\section*{Acknowledgements}
We thank Paolo Ciafaloni, Alfredo Urbano, and Ray Volkas for
helpful discussions.  NFB was supported by the Australian Research
Council, AJG and TDJ were supported by the Commonwealth of Australia,
and TJW was supported in part by U.S.~DoE grant DE--FG05--85ER40226.

\appendix

\section{Fierz transformed matrix elements}
\label{fierztransformed}

Upon Fierz transforming~(for standard $2\rightarrow 2$ Fierz identities, see e.g.,~\cite{IZp161-2,Taka1986}) 
the matrix elements of Eqn.(\ref{matrixa})--(\ref{matrixf}) we find
\begin{eqnarray}
\label{fierzedelements}
\mathcal{M}_{a}
&=&\frac{ig f^2}{\sqrt{2}q_1^2}\frac{1}{{t_1}-m_\eta^2}\frac{1}{2}
\Big(\bar v(k_2) \gamma_\alpha P_R u(k_1)\Big) \nonumber\\
&&\times
\Big(\bar u(p_1 )\gamma^\mu
\slashed{q_1}\gamma^\alpha  P_L v(p_2)\Big)\epsilon^Q_\mu,\label{MA1}\\
\mathcal{M}_{b}
&=&\frac{ig f^2}{\sqrt{2}q_{1}^2}\frac{1}{u_1-m_\eta^2}\frac{1}{2}\Big(\bar v(k_2)
\gamma_\alpha P_L u(k_1)\Big)\nonumber\\
&&\times\Big(\bar u(p_1 )\gamma^\mu
\slashed{q_1}\gamma^\alpha P_Lv(p_2)\Big)\epsilon^Q_\mu,\label{MB1}
\end{eqnarray}
\begin{eqnarray}
\mathcal{M}_{c}&=&\frac{-ig f^2}{\sqrt{2}q_2^2}\frac{1}{t_2-m_\eta^2}\frac{1}{2}\Big(\bar v
(k_2)\gamma_\alpha P_R u(k_1)\Big)\nonumber\\
&&\times\Big(\bar u(p_1 )
\gamma^\alpha \slashed{q_2} \gamma^\mu  P_L v(p_2)\Big)\epsilon^Q_\mu,\\
\mathcal{M}_{d} &=&\frac{-ig f^2}{\sqrt{2}q_2^2}\frac{1}{u_2-m_\eta^2}\frac{1}{2}\Big(\bar v
(k_2)\gamma_\alpha P_L u(k_1)\Big)\nonumber\\
&&\times\Big(\bar u(p_1 )
\gamma^\alpha \slashed{q_2} \gamma^\mu P_L v(p_2)\Big)\epsilon^Q_\mu,\\
\mathcal{M}_{e}&=&\frac{-igf^2}{2\sqrt{2}} \frac{(2k_1-2p_1-Q)^\mu}{(t_3'-m_\eta^2)(t_3-m_\eta^2)}
\left(\bar v(k_2)\gamma_\alpha P_R u(k_1)\right)\nonumber\\
&&\times\left(\bar u(p_1 )\gamma^\alpha P_L v(p_2)\right)\epsilon^Q_\mu,\\
\mathcal{M}_{f}&=&\frac{-igf^2}{2\sqrt{2}} \frac{(2k_2-2p_1-Q)^\mu}{(u_3'-m_\eta^2)(u_3-m_\eta^2)}
\left(\bar v(k_2)\gamma_\alpha P_L u(k_1)\right)\nonumber\\
&&\times\left(\bar u(p_1 )\gamma^\alpha P_L v(p_2)\right)\epsilon^Q_\mu.
\label{endfierzedelements}
\end{eqnarray}
%


Alternatively, we may apply a chiral version of the Fierz transform
(discussed in detail in Ref.~\cite{Bell:2010ei}).
to transform Eqs.(\ref{matrixa})-(\ref{matrixf}).
After a bit of algebra we get a pleasant factorized form for the
bilinear currents.  We show details for the first one, and then
summarize the results for current products of the other matrix
elements.

The current product in amplitude $M_a$ of Eq.(\ref{matrixa}) is
\begin{equation}
\label{currentsa}
\left(\bar{v}(k_2)P_L v(p_2)\right)\,\left(\bar{u}(p_1)\slashed{\epsilon}^QP_L\slashed{q}_1u(k_1)\right)\,.
\end{equation}
We write this current product in Takahashi notation~\cite{Taka1986}
and then use the chiral Fierz transform
to obtain
\begin{eqnarray}
[P_L]\,(\slashed{\epsilon}^QP_L\slashed{q}_1) &=&
\frac{1}{4}\,Tr[P_L\Gamma^C \slashed{\epsilon}^Q P_L\slashed{q}_1\Gamma_B\,]\,(\Gamma^B]\,[\Gamma_C\,)\nonumber\\
 &=& \frac{1}{4}\,Tr[P_L\gamma^\alpha  \slashed{\epsilon}^Q P_L\slashed{q}_1\gamma_\beta\,]
 \,(P_R\gamma^\beta]\,[P_L\gamma_\alpha\,)\,.\nonumber\\
 \end{eqnarray}
In going from the first equality to the second, we insert the only values for $\Gamma^C$ and $\Gamma_B$
allowed by the helicity projectors in the string of gamma matrices.
Finally, we may invert the sequence in the trace, and remove the Takahashi notation to write
the result as
\begin{eqnarray}
\frac{1}{4}\,&&Tr[P_R\slashed{\epsilon}^QP_L\slashed{q}_1\gamma_\beta\gamma^\alpha\,]\times\nonumber\\
&& \left(\bar{u}(p_1)P_R\gamma^\beta v(p_2)\right)\left(\bar{v}(k_2)P_L\gamma_\alpha u(k_1)\right)\,.
\end{eqnarray}

Amplitude $M_b$ is computed in a similar way.
In addition, it is useful to use the identity for a Majorana current
\begin{equation}
\left(\bar{v}(k_1)P_L\gamma_\alpha u(k_2)\right) =
\left(\bar{v}(k_2)P_R\gamma_\alpha u(k_1)\right)\quad[{\rm Majorana}\,]
\end{equation}
to put the final result in a form similar to that for amplitude $M_a$.
The other  current products are reduced in a similar fashion.
The final result for the product of currents after Fierzing is
\begin{eqnarray}
\label{FierzedFinal}
&\frac{1}{4}&\left(\bar{v}(k_2)P\gamma^\alpha u(k_1)\right)\left(\bar{u}(p_1)P_R\gamma^\beta v(p_2)\right) \\
&&\times
\left\{
\begin{array}{ll}
       Tr[P_R\slashed{\epsilon}^Q \slashed{q}_1\gamma_\beta\gamma_\alpha\,]\,,&\quad{\rm for\ }M_a,\,M_b\\
       Tr[P_L\slashed{\epsilon}^Q \slashed{q}_2\gamma_\beta\gamma_\alpha\,]\,,&\quad{\rm for\ }M_c,\,M_d\\
       2\,g_{\alpha\beta}\,,&\quad{\rm for\ }M_e,\,M_f\,.
\end{array}
\right.
\nonumber
\end{eqnarray}
In addition,
the unspecified projector $P$ in the first common factor is $P_L$ for amplitudes $M_a$, $M_c$, $M_e$, and
$P_R$ for the amplitudes $M_b$, $M_d$, $M_f$ derived from the crossed graphs.

What can we learn from this exercise?  For graphs $M_e$ and $M_f$ the
Fierzed currents are the same as in the $2\rightarrow 2$ case.  This
fact is not surprising since in these graphs the internal $W$ emission
does not perturb the form of the currents and their product.  However,
for the other four graphs with $W$ emission occurring on a fermion
leg, the form of the current product is quite different from the
$2\rightarrow 2$ case.  With $2\rightarrow 3$ scattering, the Lorentz
index of each current need not contract directly with the other.
Referring to Table 1 of Ref.~\cite{Bell:2010ei}, one sees that
unsuppressed Majorana annihilation amplitudes become possible for the
axial vector combination
$(\gamma_5\gamma^0\,]\,[\gamma_5\vec{\gamma}_T\,)$, and for the vector
      combination $(\gamma^3\,]\,[\vec{\gamma}_T\,)$, providing the
      trace post-factors in Eq.~(\ref{FierzedFinal}) do not vanish.
      These combinations are at the heart of the unsuppression which
      we have presented in this paper.
(The role of amplitudes $M_e$ and $M_f$ is to cancel gauge non-invariant contributions 
from graphs $M_a$-$M_d$.)
See also Ref.~\cite{Lindner:2010rr} for a
comprehensive discussion of enhanced/suppressed DM annihilation modes.

Also, for $m_\eta^2 >> t,\,u$, the non-current factors in amplitudes
$M_a$ and $M_b$ are the same, as are the non-current factors in
amplitudes $M_c$ and $M_d$.  Then the subtraction of one from the
other leads to a pure axial vector coupling in the Majorana current.
This in term leads to an effectively pure axial vector coupling in the
final state lepton current.  This effective axial vector-axial vector
coupling of currents was advertised earlier.  However, for values of
$t$ and $u$ which are non-negligible when compared to $m_\eta^2$,
there is some residual vector coupling.  In this more complicated
case, it is probably best to directly calculate rates without Fierzing
the currents.  Such is the course followed in the main text of this
paper.


\end{document}